\documentclass[%
mathleft,%
final,%
]{an}
\usepackage{graphicx}
\usepackage{times}
\begin{document}

\Pagespan{1}{}
\Yearpublication{2014}%
\Yearsubmission{2013}%
\Month{1}%
\Volume{335}%
\Issue{1}%
\DOI{This.is/not.aDOI}%

\title{Deep high spectral resolution spectroscopy and chemical composition of ionized nebulae}

\author{C. Esteban\inst{1,2}\fnmsep\thanks{Corresponding author.
  \email{cel@iac.es}}
J. Garc{\'{\i}}a-Rojas\inst{1,2}, A. Mesa-Delgado\inst{3}, \and L. Toribio San Cipriano\inst{1,2}  
}
\titlerunning{High resolution spectroscopy of ionized nebulae}
\authorrunning{C. Esteban et al.}
\institute{
Instituto de Astrof{\'{\i}}sica de Canarias, c/V{\'{\i}}a L{\'a}ctea s/n, E-38200 La Laguna, Tenerife, Spain
\and
Departamento de Astrof{\'{\i}}sica, Universidad de La Laguna, E-38206 La Laguna, Tenerife, Spain
\and
Instituto de Astrof{\'{\i}}sica, Facultad de F{\'{\i}}sica, Pontificia Universidad Cat\'olica 
de Chile, Av. Vicu{\~n}a Mackenna 4860, 782-0436 Macul, Santiago, Chile}

\received{XXXX}
\accepted{XXXX}
\publonline{XXXX}


\abstract{%
  High spectral resolution spectroscopy has proved to be very useful for the advancement of chemical abundances 
  studies in photoionized nebulae, such as H\thinspace{\sc ii} regions and planetary nebulae (PNe). Classical 
  analyses make use of the intensity of bright  collisionally excited lines (CELs), which have a strong dependence on the electron 
  temperature and density. By using high resolution spectrophotometric data, our group has led the determination of 
  chemical abundances of some heavy element ions, mainly O$^{++}$, O$^{+}$ and C$^{++}$ from faint recombination lines 
  (RLs), allowing us to deblend them from other nearby emission lines or sky features. The importance of these lines 
  is that their emissivity depends weakly on the temperature and density structure of the gas. The unresolved 
  issue in this field is that recombination lines of heavy element ions give abundances that are about 2-3 times 
  higher than those derived from CELs $-$in H\thinspace{\sc ii} regions$-$ for the same ion, and can even be a factor of 70 times higher 
  in some PNe. This uncertainty puts into doubt the validity of face values of metallicity that we use as representative 
  not only for ionized nebulae in the Local Universe, but also for star-forming dwarf and spiral galaxies at 
  different redshifts. Additionally, high-resolution data can allow us to detect and deblend faint lines of 
  neutron capture element ions in PNe. This information would introduce further restrictions to 
  evolution models of AGBs and would help to quantify the chemical enrichment in s-elements produced by 
  low and intermediate mass stars. The availability of an echelle spectrograph at the E-ELT will be of paramount 
  interest to: (a) extend the studies of heavy-element recombination lines to low metallicity objects, (b) 
  to extend abundance determinations of s-elements to planetary nebulae in the extragalactic domain and to 
  bright Galactic and extragalactic H\thinspace{\sc ii} regions. }

\maketitle

\section{Hundreds of emission lines}

\begin{figure*}
\includegraphics[angle=0, width=\linewidth]{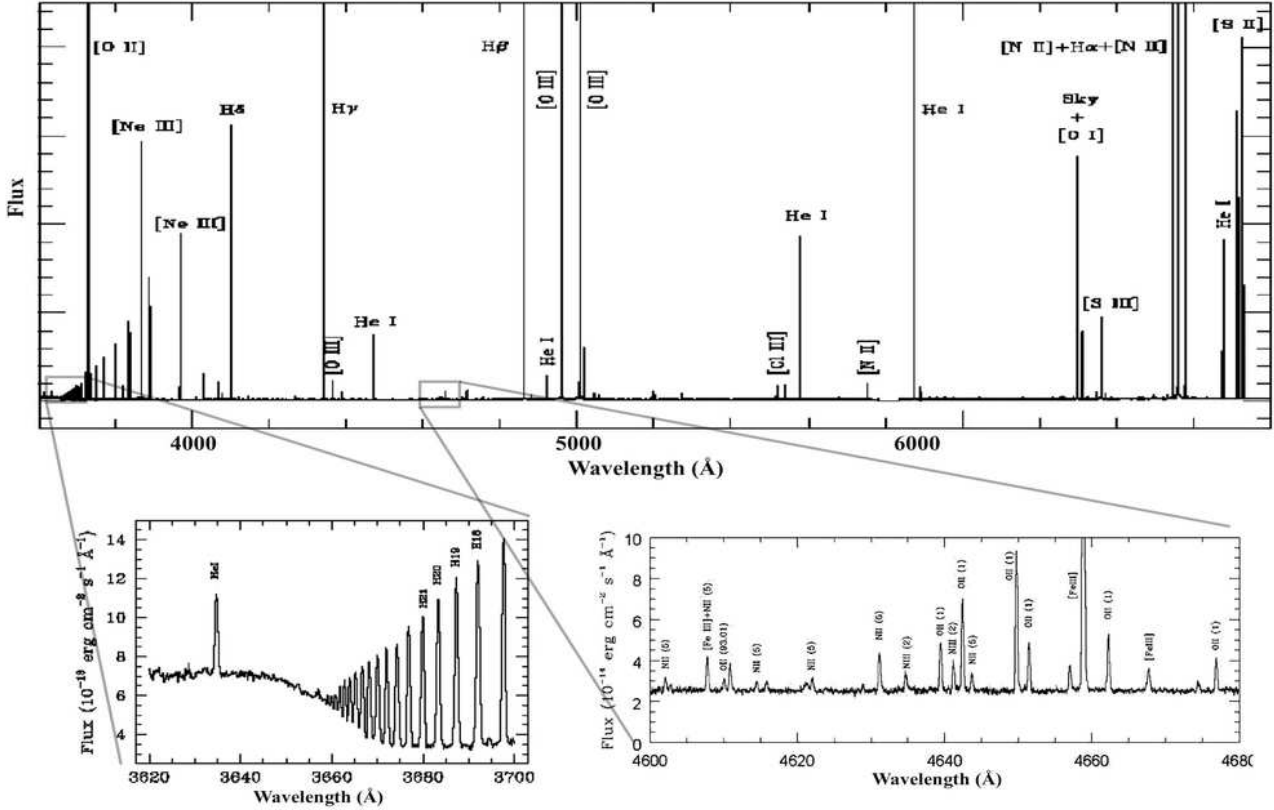}
\caption{Upper panel: optical portion of an echelle UVES@VLT spectrum of the Orion Nebula containing more than 550 emission lines, the most complete spectrum of an H\thinspace{\sc ii} region ever analyzed. The lower panels correspond to zooms of two small zones of the spectrum showing the Balmer jump and lines of high principal quantum number ($n$) of the Balmer series of H\thinspace{\sc i} (left) and the pure recombination lines of multiplet 1 of O\thinspace{\sc ii} around 4500 \AA\  (right). Data from Esteban et al. (2004) .}
\label{orion_spectrum}
\end{figure*}

High resolution spectroscopical data of ionized nebulae have been mainly used for the analysis of 
the velocity structure of the gas. Kinematics and expansion velocities of planetary nebulae (PNe, e.g. L\'opez et al. 2012) or 
wind-blown bubbles around massive stars (e.g. Esteban \& Rosado 1995) as well as turbulence in H\thinspace{\sc ii} regions 
(see Lagrois et al. 2011, and references therein) have been some of the most recurrent topics. Spectral resolutions of the order of $R$$\sim$1000 to 
3000 are commonly used for abundance determinations in ionized nebulae because those resolutions are normally sufficient to separate important emission lines 
as $-$for example$-$ the density indicator [S\thinspace{\sc ii}] doublet at 6717 and 6731 \AA\ or the temperature sensitive 
[O\thinspace{\sc iii}] 4363 \AA\ line from the mercury night-sky Hg\thinspace{\sc i} 4358 \AA\ emission line or other 
nebular lines as [Fe\thinspace{\sc ii}] 4358 \AA. However, 
high spectral resolution spectroscopy permits to obtain proper measurements of very faint lines of much interest for abundance studies due to their closeness to other lines or telluric emission features, specially in the reddest part of the optical range 
or the near infrared (NIR). In most PNe, high spectral resolution is necessary in order to resolve their usually complex velocity structure, where velocity splitting of the order of few to several tens of km s$^{-1}$ is the common rule. Since the velocity structure of H\thinspace{\sc ii} regions is normally 
much simpler, spectral resolutions of about  $R$$\sim$10,000 are necessary to resolve their line widths and thus optimize the 
contrast of the faint lines with respect to the underlying continuum. The continuum may be very intense in the case of the spectra of extragalactic 
H\thinspace{\sc ii} regions, where most of the contribution comes from the stellar underlying population we are including in our observing area and 
may even contain absorption features.  Moreover, there are also important bright collisionally excited emission lines that need the use of high spectral resolution $-$$R$$\sim$3000 or even higher$-$ for a proper deblending. These are, for example, the density indicator doublet of [O\thinspace{\sc ii}] 3726, 3729 \AA, the lines of [Ar\thinspace{\sc iv}] 4711 \AA\ and He\thinspace{\sc i} 4713 \AA\ or the [O\thinspace{\sc ii}] multiplet at about 7325 \AA\ that are very close to telluric emission features. All those lines are relevant for deriving the physical conditions of the gas. 

During the last years, our group has developed a research project mainly devoted to measure very faint pure recombination lines (RLs) of heavy-element ions in H\thinspace{\sc ii} regions and PNe from echelle spectroscopical data. The optical-NIR spectral range of the spectra of H\thinspace{\sc ii} regions contains RLs of O\thinspace{\sc i}, O\thinspace{\sc ii}, C\thinspace{\sc ii} and Ne\thinspace{\sc ii} $-$ with intensities of the order of 0.0001 to 0.001 times that  of H$\beta$$-$ that can be detected and measured in bright nebulae using large aperture telescopes. The need of  high resolution spectroscopy to detect and measure these lines is illustrated in figures 1, 2 and 3. We have obtained data for a number of bright Galactic H\thinspace{\sc ii} regions covering galactocentric distances from 6 to 12.5 kpc (see a compilation of results in Garc\'\i a-Rojas  \& Esteban 2007 and the last ones in Esteban et al. 2013) as well as a number of giant extragalactic  H\thinspace{\sc ii} regions and H\thinspace{\sc ii} galaxies (see Esteban et al. 2009 and L\'opez-S\'anchez et al. 2007). One of our most important papers is devoted to the Orion Nebula (Esteban et al. 2004) that presents line intensity ratios and a detailed analysis of a spectrum with 555 emission lines in the wavelength range from 3100 to 10,400 \AA, the most complete and deepest spectrum of an H\thinspace{\sc ii} region ever studied (see Fig.~\ref{orion_spectrum}). 

\begin{figure}
\includegraphics[angle=0, width=\linewidth]{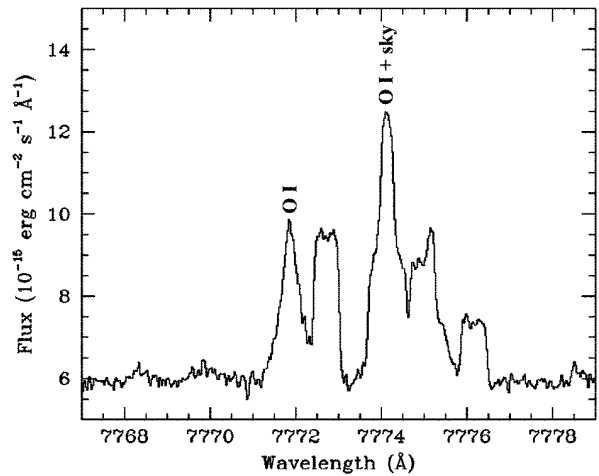}
\caption{Portion of an echelle spectrum showing the pure recombination lines of multiplet 1 of O\thinspace{\sc i} around 7770 \AA\ $-$heavily affected by telluric emission lines$-$ in the Galactic H\thinspace{\sc ii} region M8 and observed with UVES@VLT (data from Garc\'{\i}a-Rojas et al. 2007). Resolutions of the order or higher than R$\sim$8000 are needed to obtain good measurements of these lines. The flat-topped emission features correspond to sky emission.}
\label{OI1}
\end{figure}

\begin{figure}
\includegraphics[angle=0, width=\linewidth]{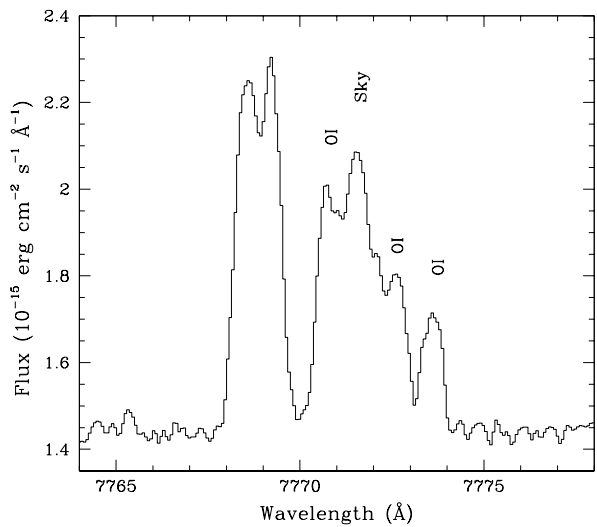}
\caption{Same as Figure 2 for the Galactic planetary nebula M\thinspace1-30 and observed with MIKE$@$Magellan at an spectral resolution of R$\sim$28,000 (Garc\'{\i}a-Rojas et al. 2012).}
\label{OI2}
\end{figure}

\section{Recombination lines and the abundance discrepancy problem}

Oxygen abundance determined from nebular spectra is the most widely used proxy of metallicity for galaxies at different redshifts. It is clear that a proper knowledge of the physical conditions and excitation mechanisms that produce emission lines in nebulae is crucial for obtaining reliable and accurate metallicities. For over three decades it has turned out that abundances of heavy-element ions determined from the standard method based on intensity ratios of collisionally excited lines (CELs) are systematically lower than those derived from the faint  RLs emitted by the same ions. This abundance discrepancy (AD) problem is far from negligible in H\thinspace{\sc ii} regions and can be very high in some PNe. In Galactic and extragalactic H\thinspace{\sc ii}  regions the O$^{2+}$/H$^+$ ratio calculated from O\thinspace{\sc ii} RLs is between 0.10 and 0.35 dex higher than that obtained from [O\thinspace{\sc iii}] CELs 
(e.g. Garc\'{\i}a-Rojas \& Esteban 2007; L\'opez-S\'anchez et al. 2007; Mesa-Delgado et al. 2009a,b; Esteban et al. 2009). Similar discrepancies have been reported for other ions with optical RLs: C$^{2+}$, Ne$^{2+}$, and O$^+$ (see Garc\'{\i}a-Rojas \& Esteban 2007), although their data are more limited and difficult to obtain. Changes or uncertainties in the face value of metallicity of celestial bodies may have a major impact in many fields of Astrophysics as the ingredients of chemical evolution models and predicted stellar yields (Carigi et al. 2005),  the luminosity --and mass-- metallicity relations for local and high-redshift star-forming galaxies (Tremonti et al. 2004), the calibration of strong-line methods for deriving the abundance scale of extragalactic H\thinspace{\sc ii}  regions and star-forming galaxies at different redshifts (Pe\~na-Guerrero et al. 2012, L\'opez-S\'anchez et al. 2012) or the determination of the primordial helium (Peimbert 2008), among others.

What is the origin of such discrepancy? Which lines shall we trust: RLs or CELs? These two fundamental questions remain unanswered but several hypotheses have been proposed in the case of H\thinspace{\sc ii}  regions. For PNe nebulae the situation is more complex because we can have material with different physical conditions and composition (see Peimbert \& Peimbert 2006). The first mechanism proposed to explain the abundance discrepancy assumes that the AD is produced by electron temperature, $T_e$, fluctuations as proposed by Peimbert (1967) (see Peimbert \& Peimbert 2011 for a recent reassessment). However, these temperature variations cannot be reproduced by photoionization models and different mechanisms have been invoked to explain their presence in ionized nebulae (see Esteban 2002; Peimbert \& Peimbert 2006). According to this scenario, RLs should provide the true abundances because their emissivities are much less dependent on temperature that those of CELs. Nicholls et al. (2012) have proposed that a $\kappa$-distribution of electron energies --departure from the Maxwell-Boltzmann distribution-- can be operating in nebulae and produce such discrepancy, being also the determinations based on RLs the more reliable ones.  A third hypothesis has been proposed by 
Stasi\'nska et al. (2007) and is based on the existence of semi-ionized clumps embedded in the ambient gas of H\thinspace{\sc ii} regions. These hypothetical clumps --unmixed supernova ejecta-- would be strong RL emitters, denser, cooler and more metallic than the ambient gas. Assuming  this scenario, abundances derived from RLs and CELs should be upper and lower limits, respectively, to the true ones, though those from CELs should be more reliable. There is even a last scenario outlined by Tsamis et al. (2011), that assumes also the presence of high-density clumps but without abundance contrast between the components. It must be hightlighted that until the AD problem is properly solved, metallicity data for 
H\thinspace{\sc ii}  regions --that come from the standard CEL method in the vast majority of cases-- should be taken with care and may even be incorrect. 

\begin{table*}
\caption{ADF for different ions and several H\thinspace{\sc ii} regions (in dex).}\label{ADFs}
\begin{tabular}{lccccccc}
\hline
Ion & Orion Neb.& 30 Dor& M8& M17& M20& NGC~3576& NGC~2579 \\
\hline
O$^+$& +0.39$\pm$0.20& +0.26$\pm$0.13& +0.14$\pm$0.09& +0.62$\pm$0.13& +0.16$\pm$0.10& $-$0.10$\pm$0.12& +0.15$\pm$0.08 \\
O$^{++}$& +0.14$\pm$0.01& +0.21$\pm$0.02& +0.37$\pm$0.04& +0.27$\pm$0.04& +0.33$\pm$0.23& +0.27$\pm$0.06& +0.27$\pm$0.04 \\
C$^{++}$& +0.40$\pm$0.15& +0.25$\pm$0.21& +0.54$\pm$0.21& ...& ...&  ...& ...\\
Ne$^{++}$& +0.26$\pm$0.10& ...& ...& ...& ...& +0.28$\pm$0.17& ...\\
Ref.& 1& 2& 3& 3& 4& 5& 6 \\
\hline
\end{tabular}
      \\
      1- Esteban et al. (2004); 2- Peimbert (2003); 3- Garc\'{\i}a-Rojas et al. (2007); 4- Garc\'{\i}a-Rojas et al. (2006); \\
      5- Garc\'{\i}a-Rojas et al. (2004); 6- Esteban et al. (2013). 
\end{table*}

Although RLs are rather faint, deep spectra taken with large aperture telescopes can provide measurements with enough signal-to-noise ratio to obtain 
precise values of the ionic abundances. In fact, the abundances obtained from RLs give very consistent results. For example, when pure RLs of more than one multiplet are observed in a nebula, the abundances obtained for each multiplet agree within the uncertainties (Liu et al. 1999; Esteban et al. 2004, 2013; Garc\'{\i}a-Rojas et al. 2004, 2007). Moreover, when the AD is measured for more than one ion for the same object, the values are quite consistent. In Table~\ref{ADFs} we present the AD factor (ADF) for the different ions ${\rm X}^{\rm i}$, defined as:
\begin{equation}
 {\rm ADF}({\rm X}^{\rm i}) = {\rm log}({\rm X}^{\rm i}/{\rm H}^+)_{\rm RLs} - {\rm log}({\rm X}^{\rm i}/{\rm H}^+)_{\rm CELs},
\end{equation} 
for the H\thinspace{\sc ii} regions where the AD has been determined for more than one ion. We can see that the ADFs are always positive (except that of O$^+$ in the case of NGC~3576, that may be affected by uncertainties of line deblending in the NIR range) and generally fairly consistent. In the case of the ADF(O$^{++}$), the mean value found for a number of Galactic and extragalactic objects is 0.26 $\pm$ 0.09 (Esteban et al. 2009). 

\section{C and O radial abundance gradients in galaxies}

Carbon is the second most abundant heavy-element in the Universe after oxygen. It is an important source of opacity and energy production in stars as well as a major constituent of interstellar dust and organic molecules. Despite its importance, we have very few determinations of its abundance in external  galaxies, mostly obtained in metal-poor dwarf irregular galaxies. The most prominent spectral features of C require observations from space. The observations of C~{\sc iii}] 1909 \AA\ and C~{\sc ii}] 2326 \AA\ lines in the UV are severely affected by uncertainties in the reddening correction. On the other hand, the far-IR [C~{\sc ii}] 158 $\mu$m fine-structure line has the clear disadvantage that its emission arises predominantly in photodissociation regions (PDRs), not in the ionized gas-phase. However, the faint RL of C~{\sc ii} 4267 \AA\ provides an alternative method to derive C abundances in ionized nebulae. Our group has been pioneer measuring the C~{\sc ii} 4267 \AA\ line in Galactic and extragalactic H\thinspace{\sc ii} regions using intermediate-high spectral resolution spectroscopy (Esteban et al. 2002, 2005, 2009, 2013; L\'opez-S\'anchez et al. 2007). In particular, we have  obtained --for the first time-- the C/H and C/O radial gradients of the ionized gas in the Milky Way (MW), and preliminary estimates for the spiral galaxies M31, M33, M101 and NGC~2403. Esteban et al. (2013) have found that the C and O abundances of NGC~2579, an H\thinspace{\sc ii} region located close to the photometric radius $R_{0}$ of the MW, show a flattening --specially evident in the C/H and C/O ratios distribution-- that can be interpreted as an effect of a leveling of the star formation efficiency at those external zones of the Galaxy. 

\begin{figure}
\includegraphics[angle=0, width=\linewidth]{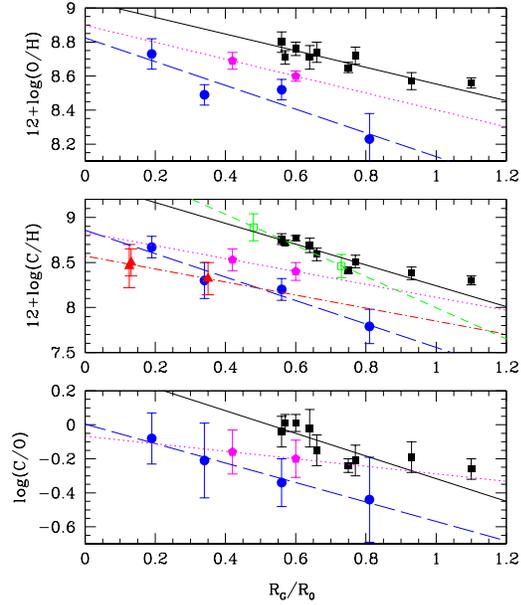}
\caption{Ionized gas phase O, C and C/O radial abundance gradients normalized to galactic photometric radius $R_0$ from H\thinspace{\sc ii} region data for the Milky Way (black squares, Esteban et al. 2005, 2013), M101 (blue circles, Esteban et al. 2009), M33 (magenta pentagons, Esteban et al. 2009), NGC~2403 (red triangles, only for C/H ratios, Esteban et al. 2009) and M31 (green empty squares, Toribio San Cipriano, in preparation). Lines represent the radial gradients of the Milky Way (black solid line, fit of all regions except the outermost one, NGC~2579, see Esteban et al. 2013), M101 (blue long-dashed line), M33 (magenta dotted line), NGC~2403 (red dashed-dotted line, only for C/H gradient) and M31 (green short-dashed 
line, only for C/H gradient). }
\label{gradients}
\end{figure}

In Fig.~\ref{gradients} we show the O/H, C/H and C/O values obtained from RLs as a function of $R_G$ normalized to their corresponding $R_0$ for 
H\thinspace{\sc ii} regions observed in the MW and the spiral galaxies M31, M33, M101 and NGC~2403. In addition, we include the corresponding radial abundance 
gradients determined from those regions. As Esteban et al. (2005) found for the MW, the C/H gradients are steeper than the ones obtained for the O/H ratio in all the galaxies producing the negative slopes of the C/O gradients. Based on chemical evolution models of the Milky Way, Carigi et al (2005), have found that 
such values of the slope of the C/O gradient can only be reproduced when considering carbon yields that increase with metallicity owing to stellar winds in massive stars.  
On the other hand, the slopes of the O/H gradients for the three galaxies represented in the upper panel of Fig.~\ref{gradients} are very similar, in contrast with the behaviour of  the C/H and C/O gradients. MW, M31 and M101 show steeper slopes than M33 and NGC~2403, which are much less massive spiral galaxies. It is clear that further deep intermediate-high resolution spectroscopical observations are needed for determining more precise values of the C/H gradients in external spiral galaxies, as well as tailored chemical evolution models of the evolution of C in those galaxies.   

Another interesting observational result for Galactic and extragalactic H\thinspace{\sc ii} regions is that the C/O ratio increases with O/H, approximately by a factor of three in the range 0.1 $<$ $Z/Z_{\odot}$ $<$ 1. At lower $Z$ the C/O ratio seems to flatten to a value that is consistent with the predictions of massive star nucleosynthesis. The main sources of C enrichment in galaxies is a matter of debate. In Carigi et al. (2005) we show that the C abundances for the MW can be explained by a combination of yields increasing with $Z$ for massive stars, and decreasing with $Z$ for low and intermediate mass stars. 

\section{C/O in planetary nebulae}

Accurate determination of C/O ratios in PNe is of paramount importance to constraint the occurrence of different nucleosynthesis processes in asymptotic giant branch (AGB) stars, as well as to put limits to the initial mass of progenitor stars of PNe by comparing with theoretical AGB evolution models. Additionally, another importance of the C/O ratio is that its value in the atmospheres of AGB stars determines the composition of the grains formed (e.g. Delgado-Inglada \& Rodr\'{\i}guez 2012). The use of RLs of C~{\sc ii}, and O~{\sc ii} to compute the C/O ratios, making the approximation C/O $\approx$ C$^{++}$/O$^{++}$  (Delgado-Inglada \& Rodr\'{\i}guez 2012) has the advantage that it avoids the problems of combining ultraviolet and optical spectra, and the use of ionization correction factors (ICFs). We have computed RL C/O ratios from deep high-resolution MIKE$@$Magellan echelle spectrophotometric data of a sample of PNe with [WC] central stars, and they have shown these type of PNe are C-richer than the average PNe (Garc\'{\i}a-Rojas et al. 2013, in preparation).

\section{Resolving the properties of gas flows}
 
\begin{figure}
\includegraphics[angle=0, width=\linewidth]{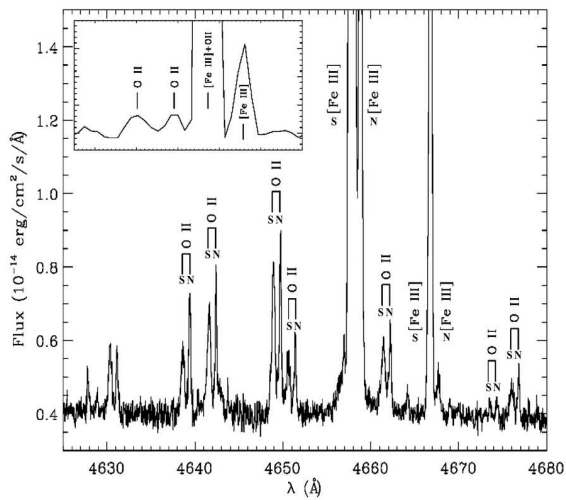}
\caption{Section of an UVES  spectrum taken with the 8m VLT of the Herbig-Haro object HH 202 in the Orion Nebula showing all the individual lines of multiplet 1 of 
O\thinspace{\sc ii} as well as some [FeIII] lines (Mesa-Delgado et al. 2009a). Note that each line shows two components, one from the ambient nebular gas (labeled as N component) and a blueshifted one corresponding to the gas belonging to the HH object itself (S or shock component). The inset on the upper left shows the same spectral zone as seen in a lower resolution spectrum, R$\sim$3000, of the same area taken with PMAS spectrograph at 3.5m CAHA telescope (Mesa-Delgado et al. 2009b).}
\label{hh202}
\end{figure}

In Mesa-Delgado et al (2009a) we present a detailed analysis of a UVES spectrum taken with the 8m VLT of the Herbig-Haro (HH) object HH~202 in the Orion Nebula. 
The spectrum shows the kinematical complexity of the gas flow at the apex of HH 202 and the spectral resolution of R$\sim$30,000 permitted to separate the shock and nebular ambient components for about 380 emission lines (see Fig.~\ref{hh202}). The relative velocity between shock and ambient components is 52 km s$^{-1}$ in this HH object. 
We were able to derive and compare accurate values of the physical conditions and chemical abundances in the shock and the ambient gas. The difference between the O$^{++}$/H$^+$ ratio determined from RLs and CELs (the abundance discrepancy factor, see Section 2) is considerably larger at the shock indicating that the gas flow has something to do with this abundance problem. The shock also shows clear evidence of dust destruction. A 30\% to 50\% of Fe and Ni atoms contained in the form of dust grains returns to the gas-phase after the passage of the shock wave. We have an ongoing program to obtain data for further photoionized HH objects in the Orion Nebula with different shock velocities and ionization conditions using UVES$@$VLT and FIES$@$NOT.
 
\section{S-elements in planetary nebulae}

\begin{figure}
\includegraphics[angle=0, width=\linewidth]{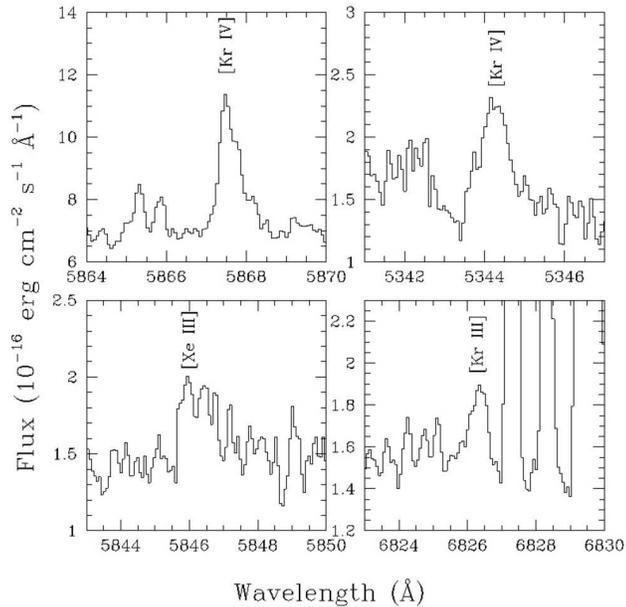}
\caption{Portion of echelle spectra showing some of the brightest collisionally excited lines of s-element ions in several planetary nebulae. The spectra have been taken with MIKE$@$Magellan at an spectral resolution of R$\sim$28,000 (Garc\'{\i}a-Rojas et al. 2012).}
\label{selements}
\end{figure}

Nearly half of the heavy elements of atomic number higher than 30 in the Universe are produced by slow neutron capture during the third dredge-up in asympthotic giant brach (AGB) stars.  
This is the so-called s-process and the elements produced are called s-elements. 
Very high spectral resolution (R$>$45,000) is mandatory in order to resolve lines of s-elements from weak features of more abundant elements (Sterling et al. 2009). Sharpee et al. (2007) carried out high-resolution observations of planetary nebulae on 4- and 6-m class telescopes, identifying lines of Br, Kr, Xe, Rb, Ba, Te, and I ions; but even at their resolution of R$\sim$22,000, many features were not unambiguously detected. Besides the observational difficulties, the spectral analysis of lines of s-elements has had the additional drawback of the scarcity of atomic data available. This situation is now changing and new calculations (e.g. Sterling \& Stancil 2011) are providing us with the appropriate tools for deriving ionic abundances as well as total ones because they are also being introduced in photoionization codes as CLOUDY. We have started an observational program to obtain deep very high resolution spectra of bright planetary nebulae in order to detect lines of s-elements ions with several echelle spectrographs (UVES$@$VLT, MIKE$@$Magellan and FIES$@$2.5mNOT; see Fig.~\ref{selements}).

\acknowledgements
This work has been funded by the Spanish Ministerio de Econom\'{\i}a y Competitividad (MINECO) under project AYA2011-22614. AMD acknowledges support from Comit\'e Mixto ESO-Chile and Basal-CATA (PFB-06/2007) grant.

%
%

\end{document}